\documentclass[pre,12pt,aps]{revtex4}
\usepackage{graphicx}
\begin{document}
\title{MIMO detection employing Markov Chain Monte Carlo}
\author{ V. Sundaram${}^{\star}$ and K.P.N.Murthy${}^{\dagger}$}
\affiliation{
${}^{\star}$ Department Electrical Engineering,\\
                          University of Notre Dame,\\
                          IN46556 Indiana, United States of America
\\
}
\affiliation{
${}^{\dagger}$ School of Physics, University of Hyderabad,\\
                            Central University P.O., Gachibowli,
                            Hyderabad 500 046, Andhra Pradesh, India 
\\
}
            
\date{\today}
\begin{abstract}
We propose a soft-output detection scheme for 
Multiple-Input-Multiple-Output (MIMO) systems. 
The detector employs Markov Chain Monte Carlo  method to compute 
bit reliabilities from the signals received 
and is thus suited for coded MIMO systems. 
It offers a good trade-off between achievable performance 
and algorithmic complexity.
\end{abstract}
\maketitle
\section{Introduction}
Multiple-Input-Multiple-Output (MIMO) systems improve 
the channel capacity many-fold by the use of multiple 
antennas at transmitter and at receiver\cite{gjf,vtnsarc}. 
It was shown \cite{gjf} that the channel capacity 
increases linearly with the number of transmit antennas 
in a rich scattering environment. Because of this 
it has become possible to 
design transmission schemes where 
a single data stream is split into several substreams 
that are simultaneously transmitted to the available 
transmit antennas. MIMO forms the core technology for 
the next generation wireless networks, as seen in 
the standards IEEE 802.11n (wireless LAN) 
as well as IEEE 802.16e (wireless MAN).

 The optimal MIMO receiver distinguishes the 
spatial signatures of different transmit substreams 
as seen at the receiver, while fully exploiting 
available receive diversity. The detection process 
also accounts for the structure of symbol 
constellations. The latter feature implies that a 
Maximum $\grave{a}$ Posterior (MAP) receiver is inherently 
more robust (in terms of error rates) to low-rank 
channels, as noted in \cite{hsadsfh}. However, the 
implementation of an exact MAP detector requires 
testing of all possible hypotheses in order to compute 
the reliability of each bit, the number of 
hypotheses being exponential in the number 
of transmit antennas and the number of bits. 
There has therefore been extensive work both 
in reducing the computational complexity of 
optimal or near-optimal detectors \cite{bhhv,jbnslbmf,eateav}
and in devising sub-optimal approaches to 
MIMO detection. Some of the latter include a 
space-time DFE with hard \cite{pwwgjf} and soft 
cancellation\cite{wjckwcjmc} and the use of iterative 
receivers \cite{bhw,sm}. The major drawback of 
all the sphere decoding algorithms and their 
variants is their worst case complexity 
(which is exponential) and the problems encountered 
in computing soft values.

Markov Chain Monte Carlo (MCMC) methods essentially consist of
 drawing  samples 
from a desired probability distribution. Multidimensional 
systems (such as MIMO) are specially suited for MCMC methods, 
whose complexity is at most polynomial with respect to 
signal dimensions. In a recent paper, Guo and Wang \cite{gw} 
proposed detection methods for MIMO systems based on sequential 
Monte Carlo (SMC) method. Dong, Wang and Doucet \cite{dwd}
applied the same 
method to a BLAST-type receiver and demonstrated that this 
can significantly improve the performance of MIMO detectors. 
Recently, Berouzhny {\it et al} \cite{fbzs} have reported 
improved performance using a Gibbs sampler. In 
this paper, we develop a Soft-In-Soft-Out (SISO) 
MIMO detector using an MCMC algorithm on a 
multidimensional lattice. 

\section{Problem formulation}

Consider a MIMO system with transmit and  receive antennas,
see Fig. (1). The source bits are encoded and then interleaved before 
being mapped to $M$ symbol streams for transmission.
Each symbol stream contains symbols drawn from the 
constellation . For a flat-fading channel $H$, 
the received signal $y$ is given by
\begin{eqnarray}
{\mathbf y}=H{\mathbf s} + {\mathbf n}
\end{eqnarray}
where ${\mathbf n}$ is the circularly symmetric channel additive 
 Gaussian white noise vector, with independent components 
each with the same variance $\sigma^2$.
For the $k^{{\rm th}}$ bit of the $m^{{\rm th}}$ symbol $s_m$, 
denoted as $b_{m,k}$, 
the MAP detector computes its $L^{{\rm th}}$ -value given by 
\begin{eqnarray}
L_{m,k}=\log\left( \frac{p(b_{m,k}=1\vert\mathbf{y})}
                        {p(b_{m,k}=0\vert\mathbf{y})}
            \right)
\end{eqnarray}  
\begin{figure}{hp}
\centering
\includegraphics[height=16mm,width=160mm]{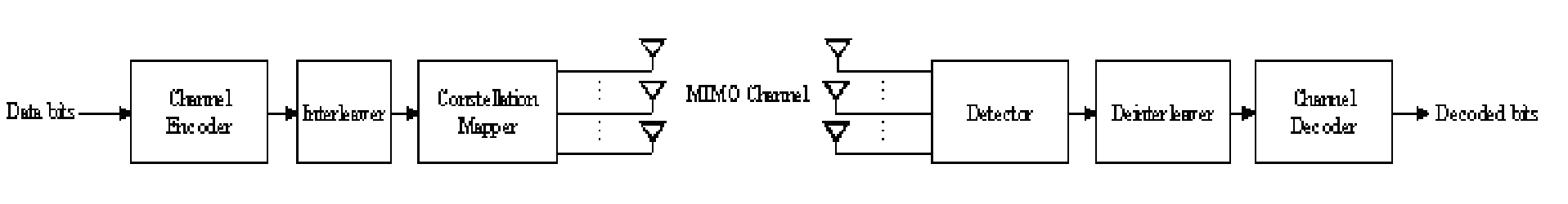}
\caption{MIMO system with transmit and receive antennas}
\end{figure}
							
\section{Markov Chain Monte Carlo Methods}

Markov Chain Monte Carlo (MCMC) methods are a family of statistical 
simulation algorithms that help construct an ensemle of
realizations or states from the desired multivariate probability 
distribution. In a typical MCMC method we start from an arbitrry
state and construct a Makov chain whose asymptotic part contains
states that belong to the desired distribution. 
This is ensured by suitably constructing a Markov transition
matrix so that its stationary distribution coincides with the 
desired distribution. With easy availability of 
high performance computing
machines in  recent times  this method has  
become popular and is being increasingly employed in a variety
of fields that include physics, chemistry, biology, economics and 
engineering see {\it e.g.} \cite{MC_ref}. 

In this paper, we use MCMC method to solve 
a Bayesian inference problem.  To this end, we identify 
the desired probability distribution (say $\rho (x)$ of the random 
variable $x$) over which the inference is to be made. 
All possible discrete values that $x$ can take, constitute the state space.
We construct a Markov chain of states.
The 
transition probability of the Markov chain determines 
the trajectory of the Markov Chain in the state space.
Asymptoticlly the Markov chin converges to the desired disribution. 
The effectiveness of MCMC metod relies on the following:
\begin{itemize}
\item[a.]
the Markov Chain quickly reaches its stationary distribution
and
\item[b.]
the desired inference parameter usually depends only 
on the most probable values of the random variable, 
that is, those states that are most frequently  visited by the Markov Chain.
\end{itemize}
Thus, spanning the entire state space that is 
usually exponential with the size of x is not 
necessary to extract useful information about the 
distribution.

\section{The Metropolis algorithm}
The Metropolis algorithm \cite{Metropolis} draws samples 
from a probability distribution $\rho (x)$, referred to as the 
target distribution in this paper. The idea behind this 
algorithm is to construct a Markov chain whose stationary 
distribution matches with the target distribution. 
Given the current state $x$
 a Metropolis sampler draws  
a candidate state, also called trial state $x_t$.
The next state in the Markov chain can be either $x$ itself or 
the trial state $x_t$. The choice is made randomly by drawing random
numbers. To this end we define a Metropolis acceptance probabiity
given by $p={\rm min}(1,\rho(x)/\rho(x_t))$. If the random number 
drawn is less than $p$ the trial state is accepted as the next entry
in the Markov chain. Otherwise the current state continues and forms the 
next state. 
The transition probability matrix $Q$ defined by 
the Metropolis algorithm,
satisfies detailed balance 
condition, given by
\begin{eqnarray}
\rho(x_i)Q(x_j\vert x_i)=\rho(x_j)Q(x_i\vert x_j)\ \ \ \forall\ \  i,j
\end{eqnarray} 									
Note that the knowledge of the trget distribution is required only upto a 
normalization  constant, which makes this algorithm 
very useful for simulating equilibrium statistical physics systems
where the normalizing partition function is not known.

\section{Bit reliability estimation using MCMC} 
The posterior probability that $b_{m,k}=1$ is givn by
\begin{eqnarray}\label{eq_posterior}
p(b_{m,k}=1\vert {\bf y})=\sum_{S_m\in S_k^{(1)}} p(S_{m,k}\vert {\bf y})
\end{eqnarray}
Here $S_k^{(1)}$ denotes all the symbols from the 
constellation $S$ that have $‘1’$ in their $k^{{\rm th}}$ bit
            position. Expressing \ref{eq_posterior} as the marginalized 
probability over all other symbols
$$S^{(-m)}=[S_0\  S_1\  \cdots\  S_{m-1}\  S_{m+1}\ \cdots]\ $$  
and applying Bayes 
theorem with prior symbol distribution
\begin{eqnarray}\label{eq_prior}
p(b_{m,k}=1\vert {\bf y})&=&
\sum_{S_m \in S_k^{(1)}} \sum_{S^{(-m)}}
                          p(S_m,{\bf S}^{(-m)}\vert {\bf y})
                                   \nonumber\\
&\propto &
\sum_{S_m \in S_k^{(1)}} \sum_{S^{(-m)}}
                          p({\bf y}\vert S_m,{\bf S}^{(-m)}
                            \vert {\bf y})
                          p(S_m,{\bf S}^{(-m)})\nonumber\\
         &\propto &
\sum_{S_m \in S_k^{(1)}} \sum_{S^{(-m)}}
                          p({\bf y}\vert S_m,{\bf S}^{(-m)})\nonumber\\
       &\approx & \sum_{{\rm sinificant\ terms}}
                          p({\bf y}\vert S_m,{\bf S}^{(-m)})
\end{eqnarray}
The final expression in the above is obtained assuming the sum 
is dominated by those symbol vectors that are most probable. 
The number of these terms will be denoted by $N_s$. 
Eq. (\ref{eq_prior}) can  be interpreted as a 
Monte Carlo integral, see \cite{fbzs}. 
A similar expression can be derived for 
$p(b_{m,k}=0\vert {\bf y})$.

\section{A Random Walk Metropolis algorithm on a symbol lattice} 

The L-values may thus be computed by a search over the symbol lattice. 
In a recent paper \cite{fbzs}, a uniform Gibbs sampler is used 
to perform the search. The algorithm may be improved by 
confining  the search to those lattice points that 
are more probable. This is captured by the likelihood 
$p({\bf y}\vert {\bf S})$ 
that can be sampled using a Metropolis algorithm.

Consider a random walk Metropolis sampler that has the 
target distribution 
$p({\bf y}\vert {\bf S})$. 
Observing 
that each vector ${\bf S}$  has a total of  $4M$ neighbors
(see Fig. 2),

\begin{figure}[hp]
\centering
\includegraphics[height=60mm,width=90mm]{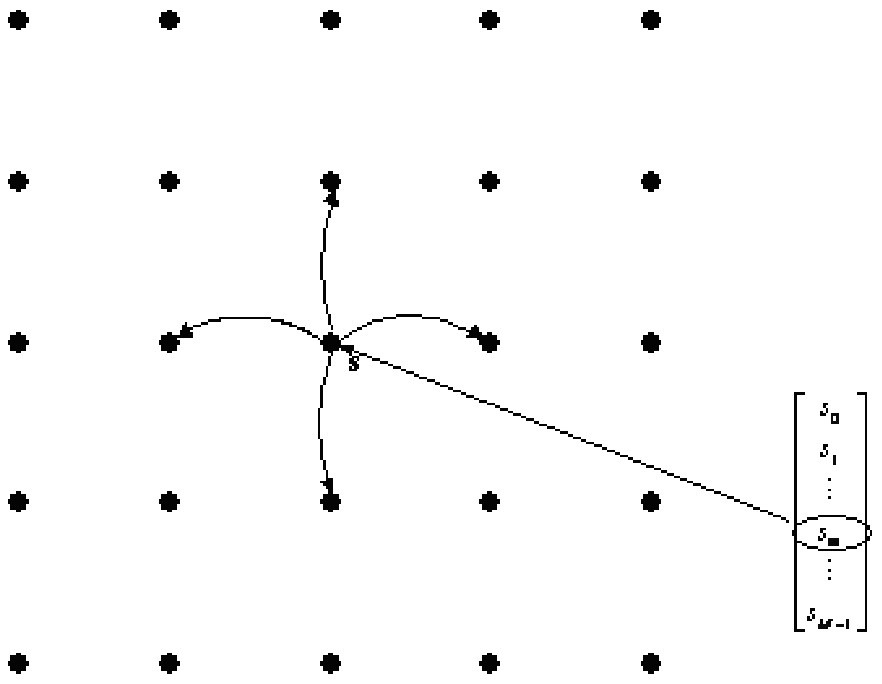}
\caption{Example of a nearest neighbour jump}
\end{figure}
\begin{eqnarray}
Q({\bf S'}\vert{\bf S})=\left\{ \begin{array}{ll}
\displaystyle\frac{1}{4M} & {\rm if}\  {\bf S'}\  {\rm is \ a\ neighbour\  of\ }
                     {\bf S}\\
                                           &\\
               0 & {\rm otherwise}\\ \end{array}\right.
\end{eqnarray}
A periodic boundary condition is applied at the lattice edges. 
At each iteration step, a trial state is randomly chosen 
amongst  the nearest neighbors of ${\bf S}$, and is accepted as the next 
state with the Metropolis acceptance probability $$p={\rm min}\left( 1,\frac{p({\bf y}\vert {\bf S'})}{
p({\bf y}\vert {\bf S})}\right)$$

\section{Simulation Results}

We now present some simulation results on a 
$3\times 3$ MIMO system, {\it i.e.}, with $3$ transmit and 
$3$ receive antennas. A packet of $64$ bytes was 
encoded using a rate-$1/2$ convolutional code 
with the generator matrix $G = [133,171]$ (in octal notation).  
After interleaving, the bits are mapped to 
a (gray-coded) $16$-QAM constellation. The MIMO 
channel is assumed to be i.i.d. Rayleigh fading, 
and changes with every channel use. At the receiver, 
the perfect CSI is assumed and the Metropolis sampler 
estimates the bit reliabilities (LLRs) using the Random 
Walk Metropolis Algorithm (RWMA) described above. 
The key parameters that determines the detector 
complexity is the number of Metropolis iterations 
(denoted by $R$), and the number of significant 
terms used (denoted by $N_s$). 
In all the  simulations, we have set $N_s=1$, that corresponds 
to the max-log-MAP approximation. We have also compared 
the results with uniform sampling. As the results 
in Figure 3 show, the frame error rate with the 
proposed algorithm is superior to that of uniform 
sampling, and seems to improve when $R$  is increased.

\begin{figure}
\centering
\includegraphics[height=100mm,width=150mm]{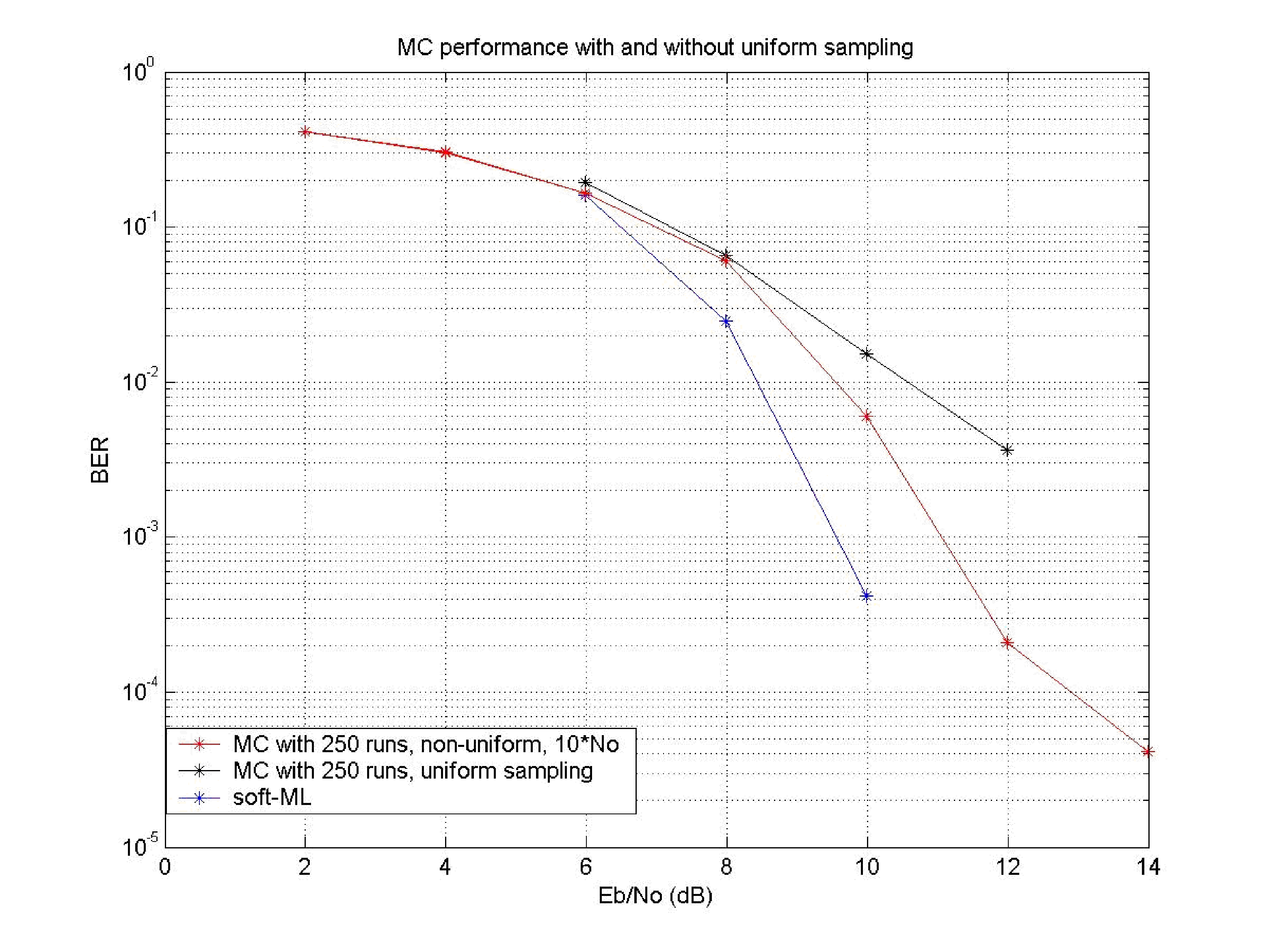}
\caption{
Simulation of RWMA algorithm on a 
3x3 16-QAM system
}
\end{figure}

The efficiency of a Metropolis sampler depends 
critically on its ability to efficiently span 
the sample space. This is captured in the 
acceptance ratio, which is defined as the 
number of number of actual state changes to 
the total number of Metropolis attempts. 
Efficient samplers have acceptance ratios 
between $0.4$ to  $0.7$. 
The argument of the exponential function 
of the prior distribution is sclaled by 
a constant referred to in statistical physics literature as
temperature. 
In several problems, the Metropolis algorithm can 
be improved by increasing the temperature. 
In 
the present case, noise plays the role of 
temperature in the system. Increasing the 
temperature improves the sample space coverage. 
In our simulations, the sampling 
temperature was set to be $10$ times that of 
the ambient noise.

\section{Conclusions}
We have proposed a soft output MIMO detector employing 
a random walk Metropolis algorithm. We find that this method
provides a good tradeoff between computational complexity 
and soft output reliability. Some interesting 
possibilities are the inclusion of a prior in the 
detection process and reducing the complexity of 
each Metropolis step. These are currently being 
investigated.

\end{document}